\documentclass[aps,prb,floatfix,twocolumn,superscriptaddress]{revtex4-2}
\usepackage{amsmath,amssymb}
\usepackage{graphicx}
\usepackage{epsfig}
\usepackage[usenames]{color}
\usepackage{verbatim}
\usepackage{psfrag}
\usepackage{xcolor}
\usepackage{hyperref}
\usepackage{dcolumn}
\usepackage{bm}
\usepackage{stmaryrd} 
\usepackage{bbm}
\usepackage{cases}

\hypersetup{colorlinks=true,citecolor={blue},linkcolor={blue},urlcolor={blue}}

\begin{document}

\title{Valley spin-acoustic resonance in MoS${_2}$ monolayers}

\author{K.~Sonowal}
\affiliation{Center for Theoretical Physics of Complex Systems, Institute for Basic Science (IBS), Daejeon 34126, Korea}
\affiliation{Basic Science Program, Korea University of Science and Technology (UST), Daejeon 34113, Korea}

\author{D.~V.~Boev}
\affiliation{Rzhanov Institute of Semiconductor Physics, Siberian Branch of Russian Academy of Sciences, Novosibirsk 630090, Russia}

\author{A.~V.~Kalameitsev}
\affiliation{Rzhanov Institute of Semiconductor Physics, Siberian Branch of Russian Academy of Sciences, Novosibirsk 630090, Russia}

\author{V.~M.~Kovalev}
\affiliation{Rzhanov Institute of Semiconductor Physics, Siberian Branch of Russian Academy of Sciences, Novosibirsk 630090, Russia}
\affiliation{Novosibirsk State Technical University, Novosibirsk 630073, Russia}

\author{I.~G.~Savenko}
\affiliation{Center for Theoretical Physics of Complex Systems, Institute for Basic Science (IBS), Daejeon 34126, Korea}
\affiliation{Basic Science Program, Korea University of Science and Technology (UST), Daejeon 34113, Korea}

\date{\today}

\begin{abstract}
The band structure of a monolayer MoS$_2$ comprises of spin-split subbands, owing to the mutual presence of broken inversion symmetry and strong spin-orbit coupling. 
In the conduction band, spin-valley coupled subbands cross each other at finite momenta, and they are valley-degenerate. 
When exposed to surface acoustic waves, the emerging strain-induced effective magnetic field can give rise to spin-flip transitions between the spin-split subbands in the vicinity of subbands crossing point, resulting in the emergence of a spin-acoustic resonance and the acoustoelectric current. 
An external magnetic field breaks the valley degeneracy resulting in the valley-selective splitting of spin-acoustic resonances both in surface acoustic wave absorption and acoustoelectric current.  
\end{abstract}

\maketitle


\section{Introduction} 
Monolayer transition metal dichalcogenides (TMDs)~\cite{Mak2010, Kormanyos2013} represent direct-bandgap semiconductors with a hexagonal lattice possessing two inequivalent valleys related to each other by time reversal symmetry. 
They are at the center-stage of recent developments in condensed matter physics, as they demonstrate a variety of fascinating physical phenomena~\cite{Wu_2019,VAEprl,OurRayleigh, Pearce_2017,Mak_2016, srivastava2015,Kodo2021,Wang2015,Wang2018}. 
One such phenomenon in monolayer TMDs is the spin-valley coupling~\cite{XiaoPRL2012} arising due to the coexistence of inversion symmetry breaking and strong spin-orbit coupling (SOC). 
As a result of spin-valley coupling, a spin splitting occurs, resulting in unique band structure with the conduction and valence bands split into spin-polarised subbands with opposite spin orientation in both the valleys.

The band structures of different TMDs have been studied lately in detail using the tight-binding model and such methods as the ${\bf k}\cdot {\bf p}$ and the density functional theory~\cite{Kosmider2013, Kormanyos2015, Kormanyos2013, Liu2013,Zhu}. 
In particular, the findings reveal different nature of spin splittings in molybdenum-based compounds ($\rm MoX_2$) and tungsten-based compounds ($\rm WX_2$). 
The dispersion of spin-split electron states in MoX$_2$ possesses specific features. In particular, the spin-up and spin-down branches cross at finite value of the electron momentum, while no such crossings happen in $\rm WX_2$. 
For the particular case of MoS$_2$, an effective splitting of spin-resolved branches varies from 3 meV at small electron momenta to zero at the electron momenta corresponding to the crossing of spin-split branches.  

This property potentially allows for the emergence of electron spin-acoustic resonance (SAR) phenomenon~\cite{SARSIC} in MoS$_2$, which is the subject of this paper. 
SAR involves the selective absorption of energy by the 2D electron gas from the acoustic vibrations corresponding to energies equal to the difference between the spin-split subbands, resulting in resonant transitions between the subbands due to spin-phonon interactions. 
We calculate the transition probability of spin-flip processes induced by surface acoustic waves (SAWs) in semiconducting structure consisting of a MoS$_2$ monolayer located on the semi-infinite substrate surface on which the Rayleigh SAWs travels. 
Initially discovered in paramagentic materials~\cite{Altshuler1962, Fidler1970, Bates1970, APRge,APRmi,APRpmion1},  later SAR was also studied in silicon carbide and nitrogen vacancy centers in diamond~\cite{SARSIC, cstsar, Maity2020, MacQuarrie, Golter} by exposing them to surface acoustic waves (SAWs). 
They represent a powerful tool to study a variety of transport phenomena~\cite{JansenSAW, long2020realization, kawada2021acoustic}. 
We also study the acoustoelectric current which arises in the system as a consequence of transfer of SAW momentum to the electron subsystem accompanied by the spin-flip transition processes.

Furthermore, an external magnetic field breaking the time reversal symmetry~\cite{PhysRevB.98.054510,oiwa2019time}  results in an additional Zeeman splitting of the originally degenerate bands and thus, it allows for an additional control of the phenomenon in question. 
Indeed, the Zeeman splitting is opposite in different valleys, thus breaking the valley degeneracy of the entire system which is reflected in the transition probability and the acoustoelectric~(AE) current. 
Transport phenomena dependent on the combined effect of valley and spin degree of freedom in TMDs hold a promising future in various applications~\cite{Ominato,Peterson2021}. Acoustic resonances hold crucial significance in the development of quantum technologies and devices~\cite{JansenSAW,cstsar}. Its dependence on valley degree of freedom can bring a huge enhancement and broadening of the potential of its present applicability.






\section{Dispersion of spin-split subbands in the conduction band of MoS$_2$ monolayer}

Up to the first-order in electron momentum $\bf{p}$, the effective ${\bf k\cdot p}$ Hamiltonian for two-band model in $\rm MoS_2$ reads~\cite{Kormanyos2015,XiaoPRL2012}
\begin{figure}[t]
\centering
\includegraphics[width=0.50\textwidth]{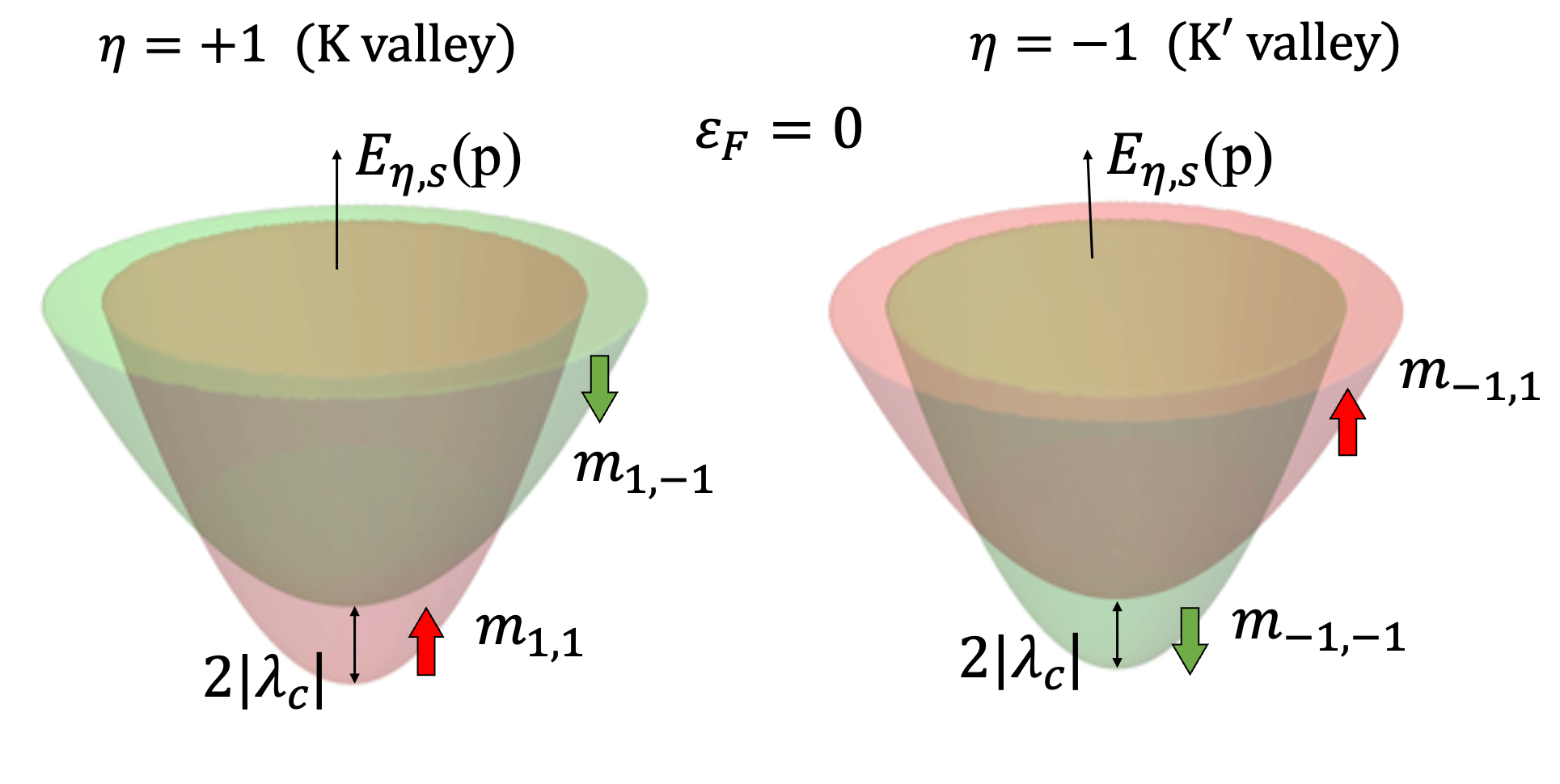}
\caption{Energy dispersions for valleys $\eta=\pm 1$ and spins $s=\pm 1$ plotted using Eq.~\eqref{massband}, where the energy is shifted by $\Delta/2$.}
\label{Fig1}
\end{figure}
\begin{equation} \label{H0}
    H_0 = v(\eta p_x \sigma_x + p_y \sigma_y) + \frac{\Delta}{2}\sigma_z,
\end{equation}
where $v$ is the Fermi velocity, $\sigma_i$ are the Pauli matrices, $\Delta$ is the gap between the conduction and the valence bands, and $\eta = \pm 1$ is the valley index.
The intrinsic SOC term describing the spin splitting of valence and conduction bands reads as~\cite{Ochoa2013}
\begin{equation}\label{Hsoc}
    H_{SO} = \frac{\lambda_c\eta}{2}(\sigma_z+\mathcal{I})s_z-\frac{\lambda_v\eta}{2}(\sigma_z-\mathcal{I})s_z,
\end{equation}
where $\mathcal{I}$ is the identity matrix in pseudospin sector, $\lambda_c$ and $\lambda_v$ are the spin-orbit splittings of the conduction and the valence bands, respectively, and $s_z$ is the spin Pauli matrix with eigenvalues $s=\pm 1$.
Thus, the total Hamiltonian reads $H = H_0 + H_{SO}$.
The eigenenergies of this Hamiltonian corresponding to the spin-split branches of the conduction band are given by (Fig.~\ref{Fig1})
\begin{gather}\label{cband}
E_{\eta,s}(\textbf{p})=\eta s\frac{\lambda_v+\lambda_c}{2}+\sqrt{\left[\frac{\Delta-\eta s(\lambda_v-\lambda_c)}{2}\right]^2+v^2p^2},
\end{gather}
and the corresponding eigenstates of the electrons in conduction band are
\begin{gather}\label{wavefunction}
\Psi_{s\eta}(\textbf{r})=\left(
                           \begin{array}{c}
                             \cos\left(\frac{\theta_{s\eta}}{2}\right) \\
                             \sin\left(\frac{\theta_{s\eta}}{2}\right)\frac{p_+}{p} \\
                           \end{array}
                        \right)\frac{e^{i\textbf{p}\cdot{\bf r}/\hbar}}{\sqrt{S}},
\end{gather}
where $p_+=\eta p_x+ip_y;\,\,p=|\textbf{p}|$, and $S$ is the monolayer area.
In Eq.~\eqref{wavefunction},
\begin{gather}\label{costheta}
\cos\theta_{s\eta}=\frac{\Delta-s\eta(\lambda_v-\lambda_c)}{2\sqrt{\left[\frac{\Delta-s\eta(\lambda_v-\lambda_c)}{2}\right]^2+v^2p^2}}.
\end{gather}
Two spin-resolved branches described by Eq.~\eqref{cband} cross if $\lambda_v>0$ and $\lambda_c<0$, as it is the case for MoS$_2$ monolayer~\cite{Kormanyos2015}.
Expanding the energy in~\eqref{cband} for small momenta $vp\ll\Delta$ gives the spin-dependent energy branches in effective-mass representation,
\begin{gather}\label{massband}
E_{\eta,s}(\textbf{p})\approx\frac{\Delta}{2}-\eta s|\lambda_c|+\frac{{\bf p}^2}{2m_{\eta,s}},\\\nonumber
\frac{1}{m_{\eta,s}}=\frac{2v^2}{\Delta-\eta s\lambda_v},
\end{gather}
where the term proportional to $\lambda_c$ was omitted in the expression for the effective mass due to its smallness, $|\lambda_c|\ll\lambda_v$ \cite{Ochoa2013}.



\section{spin-lattice interaction due to Rayleigh SAW}



\subsection{Rayleigh SAW interaction with electron spin}

Let us assume the dielectric substrate surface with MoS$_2$ monolayer correspond to the plane $z=0$ with the axis $z$ directed to the bulk of the substrate.
The monolayer is pof n-type, thus, the Fermi level lyes in the Conduciton band.
Rayleigh SAWs propagating along the substrate surface create longitudinal and transverse deformations. 
For simplicity, we assume an isotropic substrate, thus, the wave equation for Rayleigh SAWs in terms of the substrate displacement vector $\mathbf{u}$ reads~\cite{Boev2015},
\begin{equation}\label{wave}
\mathbf{\Ddot{u}} = c_t^2 \Delta \mathbf{u} + (c_l^2 - c_t^2) \rm grad~\rm div~\mathbf{u},
\end{equation}
where $c_l$ and $c_t$ are longitudinal and transverse sound velocities, respectively. 
If the SAW propagates in the $\hat{x}$ direction, the displacement vector $\bf u$({\bf r},t) has the following components~\cite{landau1986theory}:
\begin{eqnarray}\label{components}
u_x(z) &=&(kBe^{\kappa_lz} + \kappa_tAe^{\kappa_t z})e^{ikx-i\omega t},\\
\nonumber
u_y &=&0,\\
\nonumber
u_z(z) &=& (-i\kappa_lBe^{{\kappa}_lz} - ikAe^{{\kappa}_tz})e^{ikx-i\omega t},
\end{eqnarray}
where
\begin{equation}\label{kappa_lt}
\kappa_l = \sqrt{k^2 - \omega^2/c_l^2},~~ \kappa_t = \sqrt{k^2 - \omega^2/c_t^2}, ~~~~k=|\bf{k}|,
\end{equation}
and
\begin{equation}\label{coefficients}
A = \frac{\sqrt{I_0}  }{\omega \sqrt{c_t \chi\rho\kappa}},~~~~~ B = -\frac{2\sqrt{1-\xi^2}}{(2-\xi^2)}\frac{\sqrt{I_0} }{\omega \sqrt{c_t \chi\rho\kappa}},
\end{equation}
where
\begin{equation}\label{kappa}
\kappa = \frac{2(1-\chi^2)}{k_l(2-\chi^2)^2}(k_l^2 + k^2) + \frac{k^2 + k_t^2}{2k_t} - \frac{4\sqrt{1-\chi^2}k}{(2-\chi^2)}.
\end{equation}
Here $\chi$ is a constant characterising the SAW dispersion in such a way that $\omega = c_t \chi k$, $I_0$ is the SAW intensity, ${\bf k}$ is the SAW wavevector, and $\rho$ is the density of the substrate material~\cite{landau1986theory}.


The Hamiltonian capturing the effects of strain and curvature to the spin dynamics of electrons in TMDs in terms of the out-of-plane displacement $u_z$ at $z=0$ reads as~\cite{Pearce}
\begin{gather}\label{V_hamiltonian}
    V(\textbf{r},t)=\left(
    \begin{array}{cc}
      \eta \textbf{s}\cdot\textbf{B}_c(\textbf{r},t) & \beta(i\hat{s}_x+\eta \hat{s}_y)\nabla^2 u_z \\
      \beta(-i\hat{s}_x+\eta \hat{s}_y)\nabla^2 u_z & \eta \textbf{s}\cdot\textbf{B}_v(\textbf{r},t) \\
    \end{array}
  \right),
\end{gather}
where $\hat{s}_i$ are Pauli spin matrices, $\nabla^2$ is the Laplace operator, $\textbf{B}_c(\textbf{r},t)$ and $\textbf{B}_v(\textbf{r},t)$ are effective magnetic fields expressing the interaction between deformation field and the spin degree of freedom for electrons in TMD monolayers in conduction ($c$) and valence ($v$) band,
\begin{eqnarray}\label{eff_Bfield}
\textbf{B}_c(\textbf{r},t)=(2\xi^c_x\partial^2_{xy} u_z,\xi^c_y(\partial^2_x-\partial^2_y) u_z(z=0),0),\\
\nonumber
\textbf{B}_v(\textbf{r},t)=(2\xi^v_x\partial^2_{xy} u_z,\xi^v_y(\partial^2_x-\partial^2_y) u_z(z=0),0).
\end{eqnarray}
The deformation constants $\beta$ and $\xi^{c,v}_{x,y}$ describes the strength of spin-strain interaction~\cite{Pearce}. 
Thus, the interaction of a Rayleigh SAW with a $\rm MoS_2$ lattice takes place due to the coupling of spin with strain-induced effective magnetic fields. 

Substituting Eq.~\eqref{components} and Eq.~\eqref{eff_Bfield} in Eq.~\eqref{V_hamiltonian} yields
\begin{gather}\label{VR}
V(\textbf{r},t)=-\eta k^2(i \kappa_l B + i k A)\\\nonumber
\times\left(
    \begin{array}{cc}
      \xi_c\hat{s}_y & i\beta\hat{s}_- \\
      -i\beta\hat{s}_+ & \xi_v\hat{s}_y 
    \end{array}
  \right)e^{ikx-i\omega t},
\end{gather}
where $\hat{s}_\pm =\hat{s}_{x} \pm i\hat{s}_{y}$. 
%
%
%
%
%
%
Finding the conduction-band eigenfunctions, $\Psi_{s,\eta} ({\bf{p}})$ from Eq.~\eqref{wavefunction} in the limit $vp \ll \Delta$ and
using Eqs.~\eqref{VR} and~\eqref{cband} allows us to find the transition matrix element capturing the spin-lattice interaction,
\begin{gather}
M^{\eta}_{s',s}({\bf{p'}},{\bf{p}}) = \eta k^2 \xi_c \langle s'| \hat{s}_y|s\rangle(i \kappa_l B + i k A)\\
\nonumber
\times\frac{(2\pi)^2}{S}\delta\left(\frac{{\bf{p}}}{\hbar} + {\bf{k}} - \frac{{\bf{p}}'}{\hbar}\right),
\end{gather}
or,
\begin{gather}\label{Msquare}
    |M^{\eta}_{s',s}({\bf{p}}',{\bf{p}})|^2 = |M_0|^2\delta^2\left(\frac{{\bf{p}}}{\hbar} + {\bf{k}} - \frac{{\bf{p}}'}{\hbar}\right) \frac{(2\pi)^4}{S^2},
\end{gather}
where $|M_0|^2 =  k^4 \xi^2_c|\kappa_l B + k A|^2$ only depends on the parameters of the Rayleigh SAW.

\begin{figure*}
    \centering
    \includegraphics[width=0.99\textwidth]{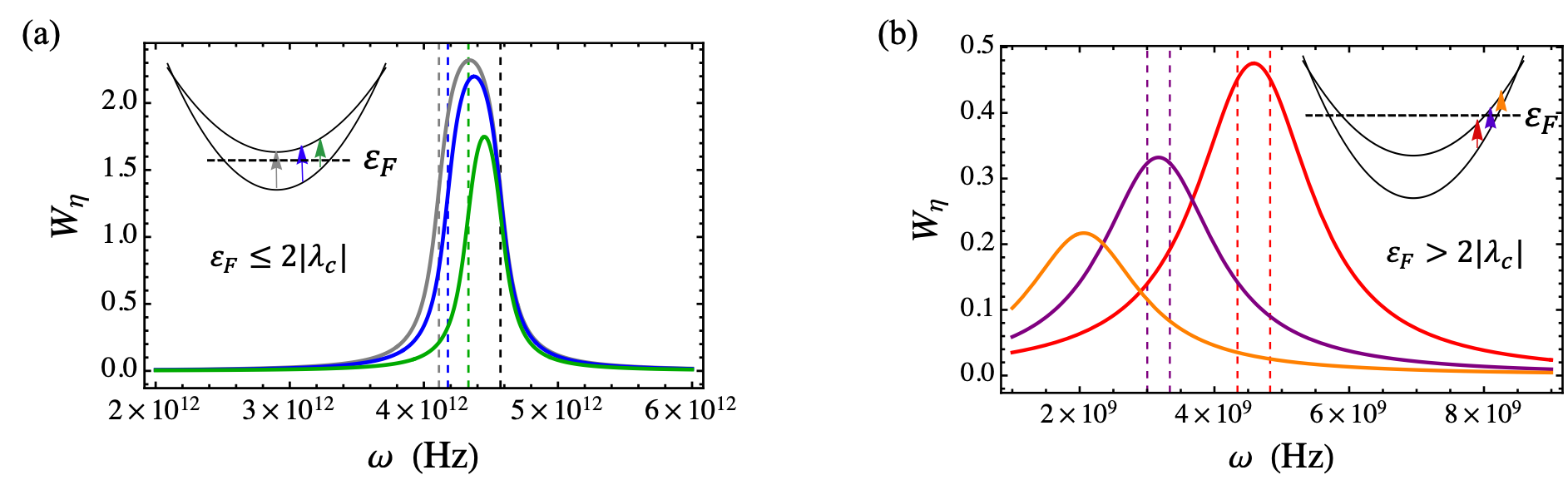}
    \caption{Normalized transition probability ${\rm W}_{\eta}/(\mu S/2\pi^2\hbar^2)$ as a function of SAW frequency.
    The vertical lines correspond to the cut-off frequencies at the onsets of resonances for $2|\lambda_c|-\varepsilon_Fm_{\eta,s}/\mu$. Offset of resonance occurs in (a) at a common frequency $2|\lambda_c|$ (black gridline) and in (b) at $(2|\lambda_c|-\frac{m_{\eta,s}}{\mu}\varepsilon_F)\frac{m_{\eta,s'}}{m_{\eta,s}}$ (gridlines of corresponding colors).The corresponding spin flip transitions are depicted in the same colors in the inset plots.}
    \label{Fig2}
\end{figure*}


\subsection{Spin acoustic resonance}

The transition rate from an initial state $|\eta,s, \bf{p}\rangle$ to the final state $|\eta,s', \bf{p'}\rangle$ in a valley $\eta $ obeys the Fermi golden rule,
\begin{eqnarray}
\nonumber
g^\eta_{s's}({\bf p}',{\bf p})&=&\frac{2\pi}{\hbar}\sum_{\bf{p}'\bf{p}}|M_{s's}^{\eta}(\textbf{p}',\textbf{p})|^2[f_{\eta s}(\textbf{p})-f_{\eta s'}({\bf p'})]\\
\label{trate}
&&\times\delta\left(E_{\eta s'}({\bf p'})-E_{\eta s}({\bf p})-\hbar\omega\right).
\end{eqnarray}
Substituting Eq.~\eqref{Msquare} in Eq.~\eqref{trate} and then, converting the summation over ${\bf p'}$ to an integration, and using the translation property of the $\delta$-function yields

\begin{gather}\label{gexpression}
g^\eta_{s's}({\bf p})=\frac{2 \pi}{\hbar} |M_0|^2 {\rm W}_{\eta},
\end{gather}
where
\begin{eqnarray}\label{W-expression}
    {\rm W}_{\eta} &=& \sum_{{\bf p}}[f_{\eta s}({\bf p})-f_{\eta s'}({\bf p}+ {\hbar\bf k})]\\
    \nonumber
 &&\times\delta\left(E_{\eta s'}({\bf p}+ {\hbar\bf k})-E_{\eta s}(\textbf{p})-\hbar\omega\right).
\end{eqnarray}
With account of electron relaxation, the delta function transforms into a Lorentzian, and then, the transition probability reads 
\begin{gather}\label{WT-expression}
{\rm W}_{\eta} =\frac{\hbar}{\pi\tau}\sum_{\bf p}  \frac{f_{\eta s}({\bf p})-f_{\eta s'}({\bf p}+{\hbar\bf k})}{\Big(E_{\eta s'}({\bf p}+{\hbar\bf k})-E_{\eta s}({\bf p}) - \hbar\omega \Big)^2 + \Big(\frac{\hbar}{\tau}\Big)^2},
\end{gather}
where, we used the phenomenological relaxation time, $\tau$, without considering the details of a particular microscopic mechanisms of electron relaxation. 

Writing the Fermi distribution functions in low-temperature limit, and substituting the energies gives 
\begin{gather}\label{W_tilde2}
    {\rm W}_{\eta}= \frac{\hbar}{\tau\pi}\sum_{{\bf p}} \Bigg[\Theta\Big(\varepsilon_F -\frac{{\bf p}^2}{2m_{\eta,s}}\Big)\\
    \nonumber
   -\Theta\Big(\varepsilon_F - \frac{({\bf p}+\hbar{\bf k})^2}{2m_{\eta,s'}}- 2|\lambda_c|\Big)\Bigg]\\
    \nonumber
    \times \frac{\hbar/\tau}{\Big( \frac{({\bf p}+\hbar{\bf k})^2}{2m_{\eta,s'}} + 2|\lambda_c| - \frac{{\bf p}^2}{2m_{\eta,s}}-\hbar\omega\Big)^2+\Big(\hbar/\tau\Big)^2}.
\end{gather}
Let us note, that the $k$-dependent terms in the denominator in the third line of Eq.~\eqref{W_tilde2} can be omitted in the limit $vk\tau \sim lk \ll 1$, where $l$ is electron mean free path. 

The second $\theta$-function in Eq.~(\ref{W_tilde2}) vanishes when $\varepsilon_F < 2|\lambda_c|$ categorising spin-flip transitions in two distinct regimes: (i) transitions, which occur when only the lower band is filled ($\varepsilon_F<2|\lambda_c|$) and (ii) when both the bands are filled ($\varepsilon_F>2|\lambda_c|$).
Performing the integration in Eq.~\eqref{W_tilde2} yields
\begin{eqnarray}\label{Wtilde}
    {\rm W}_{\eta}&=& \frac{\mu S}{2\pi^2 \hbar^2} \Bigg\{ \arctan\Big[ (\hbar\omega-2|\lambda_c| + \frac{m_{\eta,s}}{\mu}\varepsilon_F)\frac{\tau}{\hbar}\Big]  \\
    \nonumber
    &&-\arctan \Big[ (\hbar\omega-2|\lambda_c|) \frac{\tau}{\hbar}\Big]\Bigg\},
\end{eqnarray}
for $\varepsilon_F \leq 2|\lambda_c|$, and
\begin{eqnarray}\label{case2Wtilde}
    {\rm W}_{\eta} &=& \frac{\mu S}{2\pi^2\hbar^2}\Bigg\{\arctan\Big[(\hbar\omega-2|\lambda_c| + \frac{m_{\eta,s}}{\mu}\varepsilon_F)\frac{\tau}{\hbar} \Big]
    \\\nonumber
    &&-\arctan\Big[(\hbar\omega-2|\lambda_c|+\frac{m_{\eta,s'}}{\mu}(\varepsilon_F
    -2|\lambda_c|))\frac{\tau}{\hbar}\Big]\Bigg\}
\end{eqnarray}
for $\varepsilon_F > 2|\lambda_c|$, where   $\mu$ is the reduced mass $(1/\mu) = (1/m_{\eta,s})- (1/m_{\eta,s'})$, which is equal in both the valleys.
Also, note ${\rm W}_{\eta=1} = {\rm W}_{\eta=-1}$.

Figure~\ref{Fig2} shows the behavior of transition probabilities described by Eq.~\eqref{Wtilde} and Eq.~\eqref{case2Wtilde}. 
Panel (a) corresponds to Eq.~\eqref{Wtilde}, whereas panel (b) demonstrates Eq.~\eqref{case2Wtilde}. 
Adjusting the Fermi level $\varepsilon_F$ via the electron density, we obtain the curves for transition probability in Fig~\ref{Fig2}(a) for $\varepsilon_F \sim 2$ meV at $\tau=10^{-11} s$ and in Fig~\ref{Fig2}(b) $\varepsilon_F \sim 30$ meV at $\tau=10^{-9} s$. The value of spin-orbit splitting is taken as $|\lambda_c| = 1.5$ meV for MoS$_2$ monolayer~\cite{Kormanyos2015}.
In both cases, the onset of SAR occurs at SAW frequency $\omega = 2|\lambda_c|-(m_{\eta,s}/\mu)\varepsilon_F$. 
However, the upper cut-off of the frequency is different:  while in Fig.~\ref{Fig2}(a), it is always $2|\lambda_c|$, irrespective of the Fermi energy, in Fig.~\ref{Fig2}(b), the cut-off depends on the Fermi energy, $\omega = \Big(2|\lambda_c|-(m_{\eta,s}/\mu)\varepsilon_F\Big)(m_{\eta,s}/m_{\eta,s'})$. 
Thus, the SAW frequency should satisfy the condition,
\begin{gather}
    2|\lambda_c|-\frac{m_{\eta,s}}{\mu}\varepsilon_F \leq \omega \leq \Big(2|\lambda_c|- \frac{m_{\eta,s}}{\mu}\varepsilon_F\Big)\frac{m_{\eta,s}}{m_{\eta,s'}}.
\end{gather}
\begin{figure*}[htp]
    \centering
    \includegraphics[width=0.99\textwidth]{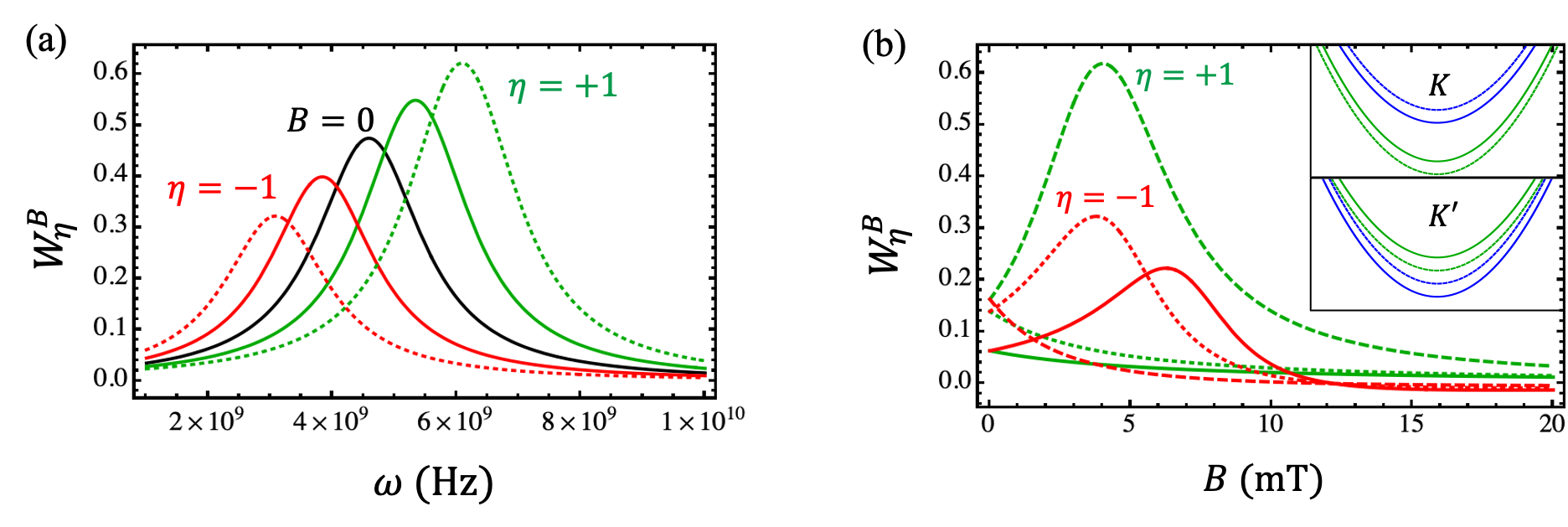}
    \caption{The illustration of the valley degeneracy breaking of the spin-split subbands in the presence of Zeeman splitting: The modified transition probability as a function of the SAW frequency (a) and the magnitude of external magnetic field (b) for the valley index $\eta = +1$ (green) and $\eta = -1$ (red). 
    Panel (a): the solid lines correspond to $B = 2$~mT, and the dotted lines correspond to $B=4$~mT. The black line corresponds to the case of zero magnetic field. 
    Panel (b): Curves of different colors correspond to different SAW frequencies: 2 GHz (solid), 3 GHz (dotted), and 6 GHz (dashed). Inset in (b) shows spin-split subbands having spin-up (green) and spin- down (blue) spins, before (solid) and after (dashed) application of external magnetic field.}
    \label{Fig3}
\end{figure*}

Panels (a) and (b) show contrasting dependence of the height and the width of the resonances on the Fermi energy: while in panel (a) ($\varepsilon_F<2|\lambda_c|$) the height and the width of each peak increases with the increase of Fermi energy (indicating an increasing probability of resonant transitions over a wider range of SAW frequencies), in panel (b) ($\varepsilon_F>2|\lambda_c|$), this probability decreases with the increase of the Fermi energy, and it ultimately goes to zero when the Fermi energy reaches the point of subbands crossing.

We note that resonant spin-flip transitions occur at SAW frequencies determined by the energy difference between the subbands (see fig.\ref{Fig2} inset).
As a consequence, SAW frequency required for resonant transition from the bottom of the band is much higher ($\sim 10^{12}$ Hz) than for transitions close to band crossings ($\sim 10^{9}$ Hz). 
Since SAW experiments usually operate with MHz to GHz frequencies, let us focus on the case $\varepsilon_F > 2|\lambda|_c$, where the Fermi level is close to the band-crossing point. 
Hence, we use Eq.~\eqref{case2Wtilde} (and not Eq.~\eqref{Wtilde}) in our further analysis.

Furthermore, the energy absorbed per unit time as a result of the spin-lattice interaction can be found after substituting ${\rm W}_{\eta}$ in Eq.~\eqref{gexpression} and multiplying this expression by a quantum of energy $\hbar\omega$, 

\begin{eqnarray}\nonumber
       {\rm Q}_{\eta} &=&  |M_0|^2   \frac{\mu S }{\pi  \hbar^3} \hbar \omega \Bigg\{\arctan\Big[(\hbar\omega-2|\lambda_c| + \frac{m_{\eta,s}}{\mu}\varepsilon_F)\frac{\tau}{\hbar} \Big]\\
    \label{resultQ}
    &-&\arctan\Big[(\hbar\omega-2|\lambda_c|+\frac{m_{\eta,s'}}{\mu}(\varepsilon_F-2|\lambda_c|))\frac{\tau}{\hbar}\Big]\Bigg\}.
\end{eqnarray}
%
Since ${\rm W}_{\eta=1} = {\rm W}_{\eta=-1}$, then ${\rm Q}_{\eta = 1} = {\rm Q}_{\eta = -1}$ also.



%
%
%

%
%
%


\subsection{Valley degeneracy breaking}
Applying an external magnetic field in $z$ direction yields a Zeeman shift of $\Delta_B=g\mu_B B$, where $g$ is the TMD monolayer electron $g-$factor and $\mu_B$ is the Bohr magneton.
The part of the Hamiltonian accounting for the spin angular momentum of electrons reads 
\begin{equation}\label{ZeemanH}
    H_B = -\Delta_B \sigma_0 s_z,
\end{equation}
and thus, the full Hamiltonian transforms into
\begin{eqnarray}\label{Htotal}
H = \begin{pmatrix}
\Delta/2 + \eta\lambda_c s -\Delta_B s & v(\eta p_x - i p_y) \\
v(\eta p_x + i p_y) &  -\Delta/2 + \eta\lambda_v s -\Delta_B s
\end{pmatrix}.~~~~
\end{eqnarray}
%
%
%
%
The energy of the spin-split subbands of the conduction band then reads,
\begin{gather}\label{energyzeeman}
E_{\eta,s}(p)\approx\frac{\Delta}{2}-\eta s|\lambda_c|+\frac{{\bf p}^2}{2m_{\eta,s}} - s\Delta_B,
\end{gather}
where a similar approximation as in Eq.~\eqref{massband} was employed.
Following the calculations as in the previous section gives the modified transition probability, ${\rm W}_{\eta}^B$ which now accounts for the external magnetic field (for $\varepsilon_F > 2|\lambda_c|$),
\begin{eqnarray}\label{WKB}
\nonumber
    {\rm W}_{\eta=+1}^{ B}&=&\frac{\mu}{2\pi\hbar^2}\Bigg\{\arctan\Big[\Big(\frac{m_{1,1}}{\mu}\varepsilon_F+\hbar\omega-2(|\lambda_c|\\
    \nonumber
   &&+\Delta_B)\Big)\frac{\tau}{\hbar} \Big]-\arctan\Big[\Big(\frac{m_{1,-1}}{\mu}(\varepsilon_F- 2(|\lambda_c|\\
    &&+ \Delta_B)) + \hbar\omega-2(|\lambda_c|+\Delta_B)\Big)\frac{\tau}{\hbar}\Big]\Bigg\}, 
\end{eqnarray}
    and
\begin{eqnarray}\label{WKBP}
\nonumber
    {\rm W}_{\eta=-1}^{ B}&=&\frac{\mu}{2\pi\hbar^2}\Bigg\{\arctan\Big[\Big(\frac{m_{-1,-1}}{\mu}\varepsilon_F+\hbar\omega-2(|\lambda_c|\\
    \nonumber
   &&-\Delta_B)\Big)\frac{\tau}{\hbar} \Big]-\arctan\Big[\Big(\frac{m_{-1,1}}{\mu}(\varepsilon_F- 2(|\lambda_c| \\
    \label{WtildeZeemanK}
    &&- \Delta_B))+ \hbar\omega-2(|\lambda_c|-\Delta_B)\Big)\frac{\tau}{\hbar}\Big]\Bigg\}.
\end{eqnarray}
Evidently, the transition probabilities differ in both the valleys. 
This difference stems from the Zeeman splittings, which are different for different valleys. 

%
%
%

%
%
%

Figure~\ref{Fig3}(a) shows the valley-dependent transition probabilities defined in Eqs.~\eqref{WKB} and \eqref{WKBP} as functions of applied SAW frequency. 
At $B=0$, the curves corresponding to different valleys overlap (black curve), and they start deviating towards opposite sides of the frequency spectrum after the magnetic field is turned on. 
This deviation increases for higher magntitudes of magnetic field. SAR peak frequency for the $K$ valley undergoes a redshift, as in this valley the distance between the spin-split subbands increases from $2|\lambda
_c| \rightarrow 2(|\lambda_c|+\Delta_B)$, thus requiring higher energy for the transition to occur. 
For similar reasons, the peak frequency in $K'$ valley undergoes a blueshift, as in this valley the inter-subband distance decreases, $2|\lambda_c| \rightarrow 2(|\lambda_c|-\Delta_B)$. 
Moreover, the lineshape of SAR remains the same for both the valleys (green and red curves). 
The deviations of these curves from the black curve (SAR curve in the valley degenerate case) are  symmetrical in each valley.

Figure~\ref{Fig3}(b) shows the dependence of spin-flip transition probabilities on magnitude of applied magnetic field at certain fixed frequencies. 
These probabilities are equal when $B=0$ (red and green curves start from the same point), and the curves start deviating from each other once $B$ is turned on. 
At some magnitudes of magnetic field (of the order of few mT), SAR peaks experience their maxima, which depend on the SAW frequency. 
Furthermore, the curves are asymmetrical. Increasing $B$ up to $\sim$20~mTs results in vanishing of ${\rm W}_{\eta}^{B}$, indicating that spin-flip transitions are suppressed. 
The energy absorbed in both valleys experiences a similar valley-dependence behavior as $\rm W_{\eta}^B$: ${\rm Q^B_K} \sim \omega {\rm W}_{\rm K}^{\rm B}$ and ${\rm Q^B_{K'}} \sim \omega {\rm W}_{\rm K'}^{\rm B}$.




\begin{figure*}
    \centering
    \includegraphics[width=0.99\textwidth]{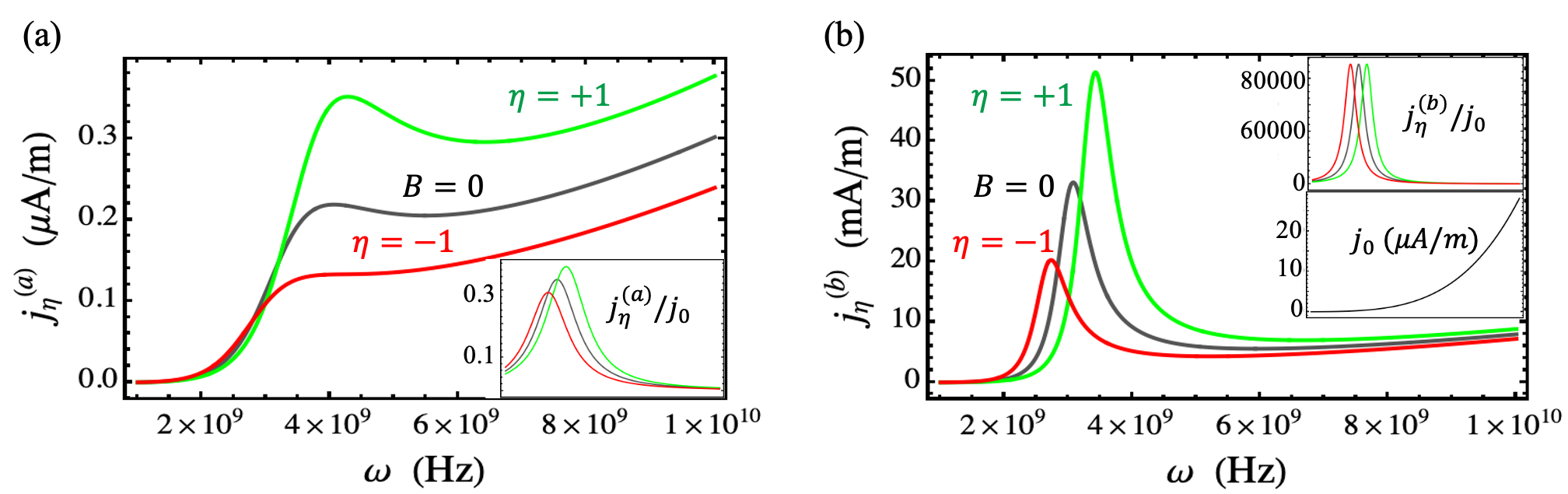}
    \caption{Two contributions of the acoustoelectric current before and after the application of external magnetic field in K and K$'$ valley as functions of frequency of the SAW. Insets in (a) and (b)(upper inset) shows the sole contribution due to valley dependent parts of the respective currents. Lower inset of (b), shows only the valley independent part, $j_0$ }
    \label{Fig4}
\end{figure*}

\section{acoustoelectric current}
The SAW-induced spin-flip transitions result in an emergence of electric current in the system due to the disbalance of electron populations in the spin-split subbands. 
The SAW transfers its momentum to the electrons under spin-flip transitions caused by the spin-SAW interaction and corresponding SAW-phonon absorption. 
The current density can be found using standard expression~\cite{Durnev},
\begin{gather}
\nonumber
    \bm{j}_{\eta} = \frac{2\pi e}{S \hbar} \sum_{{\bf p},{\bf p'}}\tau [{\bm v}^{\eta}_{{\bf p'},s'}-{\bm v}^{\eta}_{{\bf p},s}]|M^{\eta}_{s',s}({\bf p}',{\bf p})|^2 \Bigg[f\Big(E_{\eta,s}({\bf p})\Big)\\
    -f\Big(E_{\eta,s'}({\bf p}')\Big)\Bigg]\delta \Big(E_{\eta s'} ({\bf p}')-E_{\eta,s}({\bf p})-\hbar\omega \Big),
\end{gather}
where $v^{\eta}_{{\bf p},s }=dE_{\eta,s}({\bf p})/d{\bf p}$ is the electron velocity in the corresponding spin-resolved subband.

Integrating over ${\bf p'}$ in the regime $vk\tau \sim lk \ll 1$, as before, and thus, expanding the  distribution and the $\delta-$ functions for small ${\bf k}$ gives (see supplemental):

\begin{gather}
    \bm{j}_{\eta} =- C \int d{\bf p}\frac{ {\bf p}}{\mu}\Big[f\Big(E_{\eta,s}({ p})\Big)-f \Big(E_{\eta,s'}({ p})\Big)\Big]\\
    \nonumber
    \times~\hbar{\bf k}\cdot \nabla_{\bf p}\delta(E_{\eta,s'}({ p})-E_{\eta,s}({ p})-\hbar \omega)\\
    \nonumber
    +~C~\int d{\bf p} \frac{{ \bf p}}{\mu}\delta\Big(E_{\eta,s'}({ p})-E_{\eta,s}({ p})-\hbar \omega\Big)\\
    \nonumber
    \times~({\hbar\bf k}\cdot \nabla_{\bf p}f\Big(E_{\eta,s'}({ p})\Big)),    
\end{gather}
where
\begin{equation}
    C = \frac{2\pi e}{\hbar}\tau \frac{|M_0|^2}{2\pi\hbar^2}.
\end{equation} 
Accounting for relaxation processes requires the $\delta-$functions to be replaced by Lorentzians. 
In the low-temperature limit, analytical calculations give two contributions to AE current flowing in the $\hat{x}$ direction, 
\begin{eqnarray}\label{ja}
   j_{\eta}^{(a)}  =j_0\arctan \Big[\frac{
   \frac{1-\epsilon_{\lambda}}{a'}+\epsilon_{\Omega}}{\epsilon_{\tau}}
   \Big]
   -
   j_0\arctan \Big[\frac{
   \frac{1}{a}+{\epsilon_{\Omega}}}{\epsilon_{\tau}}
   \Big],~~~
\end{eqnarray}
\begin{equation}\label{jb}
    j_{\eta}^{(b)} =  -j_0\frac{2}{a}  \Bigg(1 + \frac{2}{\epsilon_T } \ln 2 \Bigg)  \frac{1/\epsilon_{\tau}}{(\frac{1}{a}+\epsilon_{\Omega})^2/\epsilon_{\tau}^2 + 1},
\end{equation}
where the SAW wavevector is asumed to be pointing ${\bf k}=(k_x,0)$, 
\begin{equation}
   j_0 = C\frac{\hbar k_x}{\mu} \frac{2\mu}{\pi},
\end{equation}
which is valley-independent, and the dimensionless parameters read
\begin{gather}
   \epsilon_F\beta=\frac{\varepsilon_F}{k_B T}= \epsilon_T,~~~~~\frac{\hbar}{\tau\epsilon_F} =\epsilon_{\tau},~~~~\frac{2|\lambda_c|}{\varepsilon_F} = \varepsilon_{\lambda},\\
    \nonumber
    \frac{\hbar \omega - 2|\lambda_c|}{\epsilon_F} = \epsilon_{\Omega},~~~~~ a'=\frac{\mu}{m_{\eta,s'}},~~~~~
    a=\frac{\mu}{m_{\eta,s}}.
\end{gather}
%

The current in the presence of magnetic field has the same form, however, $2|\lambda_c|$ entering Eqs.~\eqref{ja} and~\eqref{jb} via $\epsilon_\lambda$ should be replaced by
\begin{align}
    {\rm K~~valley:} ~~~2|\lambda_c| \longrightarrow 2(|\lambda_c| + \Delta_B),\\
    {\rm K'~~valley:} ~~~2|\lambda_c| \longrightarrow 2(|\lambda_c| - \Delta_B).    
\end{align}
Since $j_0$ does not depend on the splitting parameter $2|\lambda_c|$ originating from SOC, it is not affected by valley degeneracy breaking.

Figure~\ref{Fig4} shows two contributions to AE current. 
Both $j_{\eta}^{(a)}$ and $j_{\eta}^{(b)}$ depend on frequency through (i) the factor $j_0$ and (ii) other terms in corresponding equations. 
The factor $j_0$ is directly proportional to $\omega$ via the wavevector $k$. 
Evidently, it vanishes as $k$ goes to zero, thus, both the contributions to electric current are of drag nature. 

However, the dependence of $j_{\eta}^{(a)}$ and $j_{\eta}^{(b)}$ on frequency is more complicated than one of $j_0$, which becomes clear from the comparison of the main plots and insets in both the panels of Fig.~4.
Also, all the dependence on the valley parameter lies in (ii) terms.
%
Furthermore, it should be mentioned, that the overall magntitude of $j_{\eta}^{(b)}$ turns out orders of magnitude higher than $j_{\eta}^{(a)}$, thus, the latter can be disregarded. 

\section*{Conclusions and outlook}
In a 2D MoS$_2$ monolayer exposed to a Rayleigh surface acoustic wave there can occur a spin acoustic resonance accompanied by the onset of acoustoelectric current in the system.
%
%
Moving the doping level allows for the spin acoustic resonance to happen in two distinct frequency ranges, depending on the relation between the Fermi level $\varepsilon_F$ and the spin-orbit splitting $\lambda_c$: i) at hypersound frequencies, for $\varepsilon_F < |2\lambda_c|$, and ii) at GHz frequencies for $\varepsilon_F> 2|\lambda_c|$. 
The latter case, which occurs for electron momenta close to the subbands crossing point, seems more accessible from the experimental viewpoint due to the frequency range. 
%
The value of Fermi energy in the vicinity of crossing point of spin-resolved subbands corresponds to high electron densities. 
According to our estimations, for MoS$_2$ it is of the order of $n\approx 10^{13}-10^{14}\,cm^{-2}$. 
However, such densities are still accessible in modern experiments on TMD monolayers~\cite{costanzo2016gate,siao2018two}.

In the presence of the time-reversal symmetry in the system, the resonance and the acoustoelectric current take place in both the valleys at the same SAW frequencies. 
An application of an external magnetic field, producing different Zeeman splittings in different valleys, results in valley-dependent behavior of spin acoustic resonance and the energy absorbed by the system. 

The other possible way to lift the valley degeneracy is the illumination of the TMD monolayer sample by an external electromagnetic field. 
The specific valley-selective optical transition rules~\cite{srivastava2015, mak2012control} result in inequivalent valley populations and thus, produce different currents in different valleys even in the absence of an external magnetic field. 
In this case, the photo-induced valley spin-acoustic resonance can take place in the absence of external Zeeman field. 
Thus, the effects considered here may serve in acousto-electric spectroscopy of the valley-selective phenomena as an additional tool to monitor valley physics in TMD monolayer materials.

\section*{Acknowledgement}
We acknowledge the support by the Institute for Basic Science in Korea (Project No.~IBS-R024-D1) and the Foundation for the Advancement of Theoretical Physics and Mathematics ``BASIS''.

\bibliography{bibSAR.bib}

\end{document}